\documentclass[aps,prl,twocolumn,floatfix,showpacs]{revtex4}
\usepackage{amsmath}
\usepackage{amsfonts}
\usepackage{amssymb}
\usepackage{graphicx}

\begin{document}

\title{Spontaneous Generation of Photons in Transmission of Quantum Fields in PT Symmetric Optical Systems}
\author{G. S. Agarwal and Kenan Qu}
\affiliation{Department of Physics, Oklahoma State University, Stillwater, OK - 74078, USA}
\date{\today}

\begin{abstract}
We develop a rigorous mathematically consistent description of PT symmetric optical systems by using second quantization. We demonstrate the possibility of significant spontaneous generation of photons in PT symmetric systems. Further we show the emergence of Hanbury-Brown Twiss (HBT) correlations in spontaneous generation. We show that the spontaneous generation determines decisively the nonclassical nature of fields in PT symmetric systems. Our work can be applied to other systems like plasmonic structure where losses are compensated by gain mechanisms.
\end{abstract}
\maketitle

The PT symmetric Hamiltonians have been extensively studied in quantum mechanics~\cite{Bender1,Bender2,Bender3,Weigert,Mostafazadeh}. Generally the PT symmetric Hamiltonians involve complex potentials $V(x)$ with the property that $V(-x)=V^*(x)$. It has been well known in quantum mechanics that a complex potential implies a source or sink of energy. The well known continuity equation has a nonconservative term which is proportional to $V(x)-V^*(x)=V(x)-V(-x)$. Therefore a consistent theoretical description should account for such sources or sinks of energy. In optical systems the refractive index is like a potential~\cite{BornWolf} and since one can engineer different refractive indices, one effectively generates a very wide class of potentials leading to optical realization of PT symmetric potentials as demonstrated by Ruter et al~\cite{Ruter,Ganainy,Guo}. In optical realization of PT symmetric Hamiltonians one considers overall potential to be composed of two layers --- layer A lying in the domain say $-1<x<0$ and a layer B in the domain in $0<x<+1$. The two layers are characterized with complex optical susceptibilities $\chi$ related to complex refractive index via $n^2=1+4\pi\chi$ which are homogeneous and which have properties $\mathrm{Re}[\chi_A] = \mathrm{Re}[\chi_B]$ and $\mathrm{Im}[\chi_A] = -\mathrm{Im}[\chi_B]$. The layer A could be a gain medium and then the layer B would be an absorbing medium. Since the medium A is a gain medium it must be pumped by an external source. The medium B dissipates energy to an external environment. We must treat consistently these dissipative or gain environments, and this shows how PT symmetric optical systems can lead to significant spontaneous generation of photons. In the context of PT symmetric optical systems, the possibility of spontaneous generation, though well known in lasers, has been missed so far. This would be important when examining the transmission of single photons and entangled photons through PT symmetric structures. The transmission of single photons and in general quantum states of light through optical systems like waveguides~\cite{Politi} with $\mathrm{Im}[\chi_A] = \mathrm{Im}[\chi_B] = 0$ is currently being studied extensively, particularly in quantum information science and in problems like quantum random walks~\cite{Lobino}.

\begin{figure}[htp]
    \scalebox{0.8}{\includegraphics{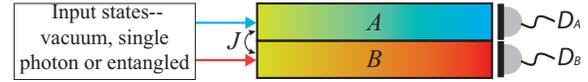}}
    \caption{\label{Fig.1} (Color online) The PT symmetric optical system consisting of two waveguides next to each other and coupled evanescently. The waveguide A(B) is filled with gain(lossy) medium. $D_A$ and $D_B$ are the detectors at the output of $A$ and $B$ for measuring intensities and intensity correlations.}
\end{figure}

Consider the PT symmetric optical system shown in Fig.~\ref{Fig.1}. The system consists of two single mode waveguides which are evanescently coupled. The waveguide A is filled with a gain medium whereas the waveguide B is filled with a lossy medium. The PT symmetry would require that the gain in A must be equal to absorption in B~\cite{note}. Let $\alpha$ and $\beta$ be the classical field amplitudes in the two waveguides. These amplitudes satisfy the basic equations
\begin{equation}\label{1}
    \frac{\mathrm{d}\alpha}{\mathrm{d}z} = G\alpha-\mathrm{i}J\beta, \quad
    \frac{\mathrm{d}\beta}{\mathrm{d}z} = -G\beta-\mathrm{i}J\alpha.
\end{equation}
We would now discuss the equations for the quantized fields in A and B. Let $a$, $b$ ($a^\dag$, $b^\dag$) be the annihilation (creation) operators for the fields in A and B. In order to obtain the quantized equations, we note that any gain or lossy medium leads to quantum fluctuations in the fields~\cite{thanks}. The basic equations can be derived from considerations of the microscopic Hamiltonian for the radiation fields, pumping and loss mechanisms as is usually done in laser theory~\cite{Scully}. We recall from Eqs. (12.1.8) and (12.1.15) that the field mode $a$ in a cavity containing an amplifying medium obeys the quantum Langevin equation $\dot a=ga+f_a$, where $2g$ is the gain and $f_a$ is a delta correlated force with zero mean value and with the only nonvanishing second order correlation $\langle f_a^\dag(t)f_a(t')\rangle = 2g \delta(t-t')$. The equations for a lossy medium are similar (Ref.~\cite{Scully} Eqs. (9.1.15) and (9.1.27) with thermal bath at zero temperature). We use the analog of these equations for the waveguides. Further, the evanescent part of the coupling is described by the Hamiltonian~\cite{Politi} $Jc(a^\dag b+ab^\dag)$. The time in the equations from laser theory is to be replaced by the propagation distance in the waveguides. For the PT symmetric optical systems, the final quantum Langevin equations are
\begin{equation}\label{2}
    \frac{\mathrm{d}a}{\mathrm{d}z} = Ga-\mathrm{i}Jb + f_a, \quad
    \frac{\mathrm{d}b}{\mathrm{d}z} = -Gb-\mathrm{i}Ja + f_b.
\end{equation}
Here $f_a$ and $f_b$ are the quantum Langevin forces with zero mean and with  the properties
\begin{equation}\label{3}
    \begin{gathered}
    \langle f_a^\dag(z)f_a(z')\rangle = 2G\delta(z-z'), \qquad \langle f_a(z)f_a^\dag(z')\rangle = 0, \\
    \langle f_b^\dag(z)f_b(z')\rangle = 0, \qquad \langle f_b(z)f_b^\dag(z')\rangle = 2G\delta(z-z'),
    \end{gathered}
\end{equation}
Further the fluctuating forces are Gaussian. Note the difference in the nature of Langevin forces in the equations for the amplifying and absorbing medium. These quantum mechanical fluctuating forces in general would contribute to the output of the waveguides and significantly modify the outputs. The quantum Langevin equations (\ref{2}) can be integrated over the length $l$ of the waveguide. We write the solution as
\begin{equation}\label{4}
    {a(l) \choose b(l)} = \mathrm{e}^{JMl} {a(0) \choose b(0)} + \int_0^l\mathrm{d}l' \mathrm{e}^{JM(l-l')} {f_a(l') \choose f_b(l')},
\end{equation}
where
\begin{equation}\label{5}
    M = \left(\begin{array}{cc}
        g  & -\mathrm{i} \\
        -\mathrm{i}  & g
        \end{array} \right),
      \qquad g=\frac{G}{J}.
\end{equation}

The matrix $M$ has the eigenvalues $\pm\mathrm{i}\sqrt{1-g^2}$ if $g=G/J<1$, $\pm\sqrt{g^2-1}$ if $g>1$. The intensities at the output $l=Jz$ are obtained by calculating $I_a = \langle a^\dag(l)a(l)\rangle$, $I_b = \langle b^\dag(l)b(l)\rangle$ for a variety of input states. The input states could be single photon states like $|1,0\rangle$, $|0,1\rangle$ and even a NOON state like $(|0,2\rangle+|2,0\rangle)/\sqrt{2}$. We next give explicit results for a variety of input quantum states in the PT symmetric optical system. Note that the biorthogonal modes of the PT symmetric optical system are given by the eigenstates of M. In view of the presence of quantum Langevin terms in Eq.(\ref{4}) the eigenmodes of the system are strongly correlated.

\section*{Spontaneous Generation of Photons in PT Symmetric Optical Systems}

\begin{figure}[h]
    \begin{center}
        \scalebox{0.35}{\includegraphics{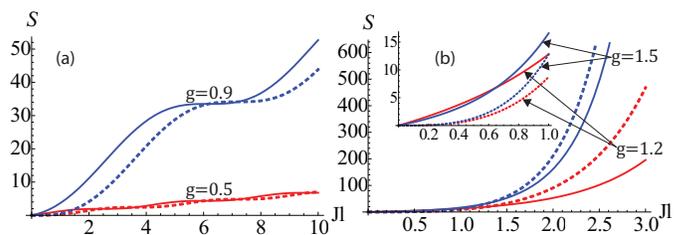}}
    \end{center}
    \caption{\label{Fig.2} (Color online) Spontaneous generation of the radiation in PT symmetric optical systems. The solid and dashed curves give respectively the output of the waveguides A and B. Notice the dashed curves are multiplied by a factor of 10. The part (a)[(b)] is for $g<1[>1]$.}
\end{figure}

Let us first consider the most intriguing possibility of spontaneous generation in PT symmetric systems. In this case the input state of photons is $|0,0\rangle$. Further the quantum noise in the absorber does not contribute to the normally ordered moments and thus only $f_a$ terms contribute to spontaneous generation. Using Eq.(\ref{3}) and (\ref{4}) we can show that the spontaneous generation, denoted by the symbol $S$, is given by
\begin{equation}\label{6}
\begin{gathered}
    S_a(l) = 2g\int_0^l\mathrm{d}l'|K_{aa}(l')|^2, \\
    S_b(l) = 2g\int_0^l\mathrm{d}l'|K_{ba}(l')|^2,
    \end{gathered}
\end{equation}
where $K_{ij}=(\mathrm{e}^{JMl})_{ij}$. We have calculated analytically the matrix $K$ and the generated signals. We show in Fig.~\ref{Fig.2}, the nature of the spontaneous generation for different values of the gain both below and above threshold $g=1$. The threshold is defined by the condition that the eigenvalues of the matrix $M$ change from pure imaginary to pure real values. We note from the Fig.~\ref{Fig.2} that the generation in the waveguide B starts later as initially the generation has to take place in the waveguide A and this must tunnel into the waveguide B to see the generation in the waveguide B. For large $g$ the generation in the waveguide A dominates over that in B
\begin{equation}\label{7}
    S_a \to \mathrm{e}^{2gl}, \qquad S_b \to \frac{1}{4g^2}\mathrm{e}^{2gl}.
\end{equation}

\section*{Vacuum Induced HBT Cross Correlation}

\begin{figure}[h]
    \begin{center}
        {\scalebox{0.35}{\includegraphics{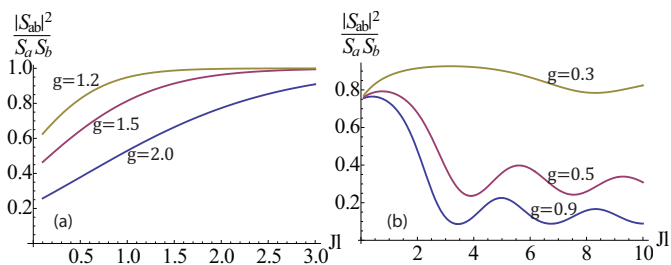}}}
    \end{center}
    \caption{\label{Fig.3} (Color online) Vacuum induced (HBT) cross correlation between the outputs of the two waveguides.}
\end{figure}

We next examine the HBT~\cite{HBT} cross correlation between the outputs of the two waveguides. We expect such a cross correlation to be nonzero as the generation in B is due to generation in A and due to tunneling effects. Let us define a correlation coefficient as
\begin{equation}\label{8}
\begin{gathered}
    q(l) = g^{(2)}-1, \\
    g^{(2)} = \frac{\langle a^\dag(l)b^\dag(l)a(l)b(l)\rangle}{\langle a^\dag(l)a(l)\rangle \langle b^\dag(l)b(l)\rangle}.
\end{gathered}
\end{equation}
The term $\langle a^\dag(l)b^\dag(l)a(l)b(l)\rangle$ is proportional to the probability of simultaneously detecting one photon each in the output of both A and B waveguides. This can be evaluated by using the Gaussian nature of the Langevin forces $f_a$ and $f_b$ which implies that the Heisenberg operators are also Gaussian with the provision that we need to keep track of the operator characteristics. This applies only to the problem of spontaneous generation. It can be shown using the Gaussian property that
\begin{multline}\label{9}
    \langle a^\dag(l)b^\dag(l)a(l)b(l) \rangle \\
    = \langle a^\dag(l)a(l)\rangle \langle b^\dag(l)b(l)\rangle + \langle a^\dag(l)b(l)\rangle \langle b^\dag(l)a(l)\rangle
\end{multline}
and hence
\begin{equation}\label{10}
    q(l) = \frac{|\langle a^\dag(l)b(l)|^2\rangle}{S_aS_b} = \frac{|S_{ab}|^2}{S_aS_b},
\end{equation}
The term $\langle a^\dag(l)b(l)\rangle$ is obtained from (\ref{4}) as
\begin{equation}\label{11}
    \langle a^\dag(l)b(l)\rangle = 2g\int_0^l\mathrm{d}l' K^*_{aa}(l')K_{ba}(l').
\end{equation}
The nature of the vacuum induced HBT cross correlation between the outputs of the two waveguides is shown in Fig.~\ref{Fig.3}. The behavior of the HBT correlation is quite different for $g<1$ and $g>1$. For $g>1$, the correlation saturates to unity, whereas for $g<1$, it tends to remain, in most cases, much smaller than unity. This is due to the distinction in the nature of the eigenvalues as discussed after Eq.(\ref{5}). Thus for large gain, the correlations or the fluctuations are maximally Gaussian in the sense that the maximum value of $g^{(2)}$ can be $2$, and hence maximum value of $q$ will be $1$.

\section*{Stimulated vs Spontaneous Generation in PT Symmetric Optical Systems}

\begin{figure}[h]
    \begin{center}
        {\scalebox{0.35}{\includegraphics{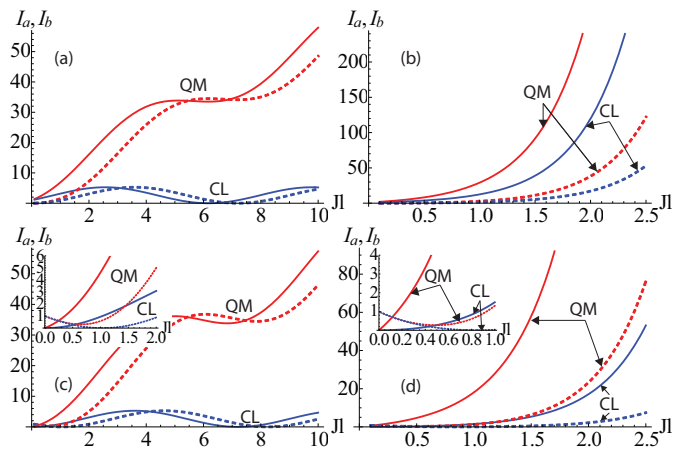}}}
    \end{center}
    \caption{\label{Fig.4} (Color online) The mean number of photons in the output of waveguides A (solid curves) and B (dashed curves) when the input state is $|1,0\rangle$ (panels a, b), $|0,1\rangle$ (panels c, d). Panels a and c are for $g=0.9$; panels b and d for $g=1.5$.}
\end{figure}

We would now consider the stimulated generation by single photon input states and compare the stimulated generation with spontaneous generation. The stimulated generation comes from the terms $\displaystyle\mathrm{e}^{Ml} {a(0) \choose b(0)}$ in Eq.(\ref{4}). For the input $|1,0\rangle$ the stimulated generation is given by
\begin{equation}\label{12}
    I_a^{st} = |K_{aa}(l)|^2, \qquad I_b^{st} = |K_{ba}(l)|^2,
\end{equation}
whereas for the input single photon in the waveguide B, the stimulated generation is given by
\begin{equation}\label{13}
    I_a^{st} = |K_{ab}(l)|^2, \qquad I_b^{st} = |K_{bb}(l)|^2.
\end{equation}
Note that for large propagation distance and $g>1$, the stimulated generation is determined by
\begin{equation}\label{14}
\begin{gathered}
    |K_{aa}(l)|^2 \to \mathrm{e}^{2gl}, \qquad |K_{bb}(l)|^2 \to \frac{1}{16g^4}\mathrm{e}^{2gl}, \\
    |K_{ab}(l)|^2 = |K_{ba}(l)|^2 \to \frac{1}{4g^2}\mathrm{e}^{2gl}.
    \end{gathered}
\end{equation}
Hence asymptotically negligibly small output is there from the waveguide B. In Fig.~\ref{Fig.4} we present explicit results for the stimulated generation and the total generation for the input single photon either in the waveguide A or in B. The stimulated part is same as from Eq.(\ref{1}) and hence labeled as "CL"(Classical) in Fig.~\ref{Fig.4}. The curves labeled as "QM"(quantum) include contributions from both stimulated and spontaneous generation. The behavior is quite different for $g<1$ and for $g>1$. For $g<1$, we find that total generation of photons is dominated by spontaneous generation. The situation is somewhat different for $g>1$. However, spontaneous generation still remains quite important.

\section*{HBT Correlations for Input Fields in a NOON State}

\begin{figure}[h]
    \begin{center}
        {\scalebox{0.35}{\includegraphics{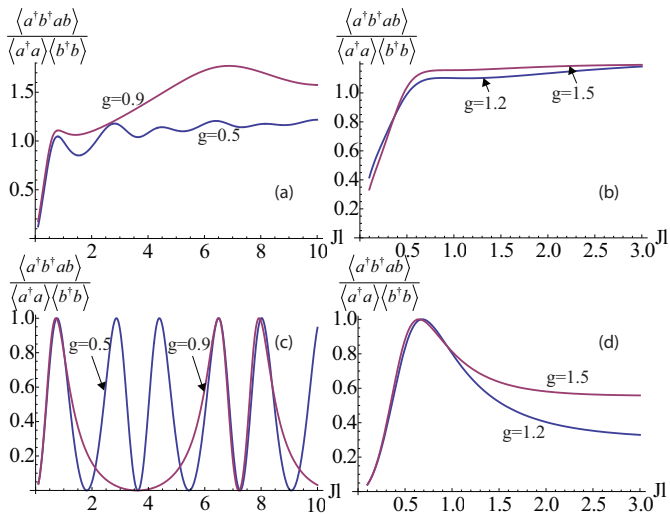}}}
    \end{center}
    \caption{\label{Fig.5} (Color online) HBT cross correlation between the two waveguides when the input is a NOON state $(|2,0\rangle+|0,2\rangle)/\sqrt{2}$. In panels a and b, we consider both spontaneous and stimulated generation; while in panels c and d, we consider only the stimulated one. }
\end{figure}

Lastly we examine the behavior of optical fields in PT symmetric systems when the input fields are in a NOON state~\cite{Dowling} $(|0,2\rangle+|2,0\rangle)/\sqrt{2}$ which is an entangled state. For this state $g^{(2)}$ has the least value which is zero. The mean intensity of emission has a behavior similar to that in case of the state $|1,0\rangle$
\begin{equation}\label{15}
\begin{gathered}
    \langle I_a\rangle = |K_{aa}|^2 + |K_{ab}|^2 +S_a, \\
    \langle I_b\rangle = |K_{ba}|^2 + |K_{bb}|^2 +S_b.
\end{gathered}
\end{equation}
The probability of finding one photon in each waveguide  is now calculated using the full solution (\ref{4})  and is found to be the sum of three different types of contributions: (i) stimulated ones $|K_{aa}K_{ba}+K_{ab}K_{bb}|^2$; (ii) purely from spontaneous emission $S_aS_b +S_{ab}S_{ba}$; (iii) cross terms involving both spontaneous and stimulated emission
\begin{multline}\label{16}
    S_a(|K_{ba}|^2+|K_{bb}|^2) + S_b(|K_{aa}|^2+|K_{ab}|^2) \\
    + \{S_{ab}K_{ba}^*K_{aa} + S_{ab}K_{bb}^*K_{ab} + c.c. \}.
\end{multline}
We show in Fig.~\ref{Fig.5} the correlation $g^{(2)}$ in case if we completely ignore spontaneous generation, i.e. the result with $S_a=S_b=S_{ab}=0$, and compare with the full result---which includes all the three contributions as mentioned above. The behavior in Fig.~\ref{Fig.5}c can be understood in terms of the eigenvalues $\pm\mathrm{i}\sqrt{1-g^2}$ of $M$ for $g<1$, as in this case we have periodic exchange of energy between the two waveguides. Fig.~\ref{Fig.5}c and Fig.~\ref{Fig.5}d show that $g^{(2)}$ is mostly less than unity if the spontaneous generation is ignored whereas the inclusion of spontaneous emission changes the character of $g^{(2)}$ completely (Fig.~\ref{Fig.5}a and Fig.~\ref{Fig.5}b). Thus noninclusion of spontaneous emission would lead to erroneous conclusion that the fields in PT symmetric system are nonclassical whereas the correct behavior leads to nonclassicality only over a small range of propagation distance.

In summary we have shown how important is the spontaneous generation of photons in PT symmetric optical systems. We show significant HBT correlations in the generated photon fields. The spontaneously generated photons strongly correlate eigenmodes and determine the quantum nature of the fields containing single and entangled photons in such systems. There are many other important systems involving for example the lossless propagation in plasmonic structures~\cite{Leon} where losses are compensated by gain media. While dealing with quantum fields in such structures one has to supplement Maxwell equations with quantum Langevin terms of the type discussed in the current work. These ideas would also be applicable, say, in the context of matter waves where a whole class of potentials including complex ones can be constructed by using off resonant laser fields~\cite{Keller}.

GSA thanks K. Babu and G. Vemuri for discussions on PT symmetric Hamiltonians and the Director Tata Institute of Fundamental Research Mumbai for hospitality during the preparation of this work.



\begin{thebibliography}{1}
\bibitem{note} The PT symmetry would be broken when the gain and loss are unequal. Such a case can be discussed using the method of the present paper.
\end{thebibliography}
\end{document}